\documentclass{emulateapj}
\usepackage{amssymb}
\usepackage{amsmath,bm}
\usepackage[normalem]{ulem}
\usepackage[dvips]{color}
\usepackage{subfigure}
\usepackage[english]{babel}
\usepackage{multirow}
\usepackage[mathcal]{eucal}
\usepackage{rotating}
\usepackage{graphicx}
\usepackage{dcolumn}
\usepackage{bm}

\renewcommand\sout{\bgroup \color{red} \ULdepth=-.5ex \ULset}

\begin{document}	

\title{Strange Quark Stars as Probe of Dark Matter}

\author{Hao Zheng$^1$, and Lie-Wen Chen$^{*1,2}$}
\affil{$^1$ Department of Physics and Astronomy and Shanghai Key Laboratory for
Particle Physics and Cosmology, Shanghai Jiao Tong University, Shanghai 200240, China}
\affil{$^2$ Center of Theoretical Nuclear Physics, National Laboratory of Heavy Ion
Accelerator, Lanzhou 730000, China}
\altaffiltext{*}{Corresponding author (email: lwchen$@$sjtu.edu.cn)}


\begin{abstract}
We demonstrate that the observation of old strange quark stars (SQSs) can set
important limits on the scattering cross sections $\sigma_q$ between the
light quarks and the non-interacting scalar dark matter (DM).
By analyzing a set of 1403 of solitary pulsarlike compact
stars in the Milky Way, we find the old solitary pulsar PSR J1801-0857D
can set the most stringent upper limits on $\sigma_q$ or the DM-proton scattering
cross sections $\sigma_p$.
By converting $\sigma_q$ into $\sigma_p$ based
on effective operator analyses, we show the resulting $\sigma_p$
limit by assuming PSR J1801-0857D to be a SQS could be comparable with that of the current direct detection
experiments but much weaker (by several orders of magnitude) than that
obtained by assuming PSR J1801-0857D to be a neutron star (NS), which requires an extremely
small $\sigma_p$ far beyond the limits of direct
detection experiments.
Our findings imply that the old pulsars are favored to be SQSs
rather than NSs if the scalar DM were observed by future
terrestrial experiments.
\end{abstract}
\maketitle

\section{Introduction}
\label{sec:introduction}
Observation of the pulsarlike old compact stars can provide constraints
on dark matter (DM). Being accreted onto the star by strong gravity,
DM can accumulate efficiently inside the star via scattering with
the star matter and eventually may collapse into a star-killing black hole (BH).
To prevent destruction of the star, the interactions
between DM and the star matter must be extremely weak.
After the pioneering work by Goldman and Nussinov~\citep{Gold89}
on this subject, many works have been devoted to constraining the
properties of DM based on the observation of old compact
stars~\citep{Ber08,Lava10,Chris11,Chris12,Yu12,Kumar13,Kumar14,Bra14,Hao15}.
Among these works,
the host compact stars are generally assumed to be neutron stars (NSs),
and thus numerous constraints on DM-nucleon interactions have
been obtained.
In particular, for bosonic DM which is favored by many theories beyond
the standard model,
if its self-interactions can be ignored,
the formation of Bose-Einstein condensate (BEC) state could further facilitate
the occurrence of DM collapsing into BH, and the resulting limits on the
DM-nucleon interactions become
far beyond the terrestrial
experiments~\citep{DAMA08,Xe12,Lux13,CDMS13,SCDMS14,Xe1t,Bill14}, leading to
a conclusion that the bosonic DM cannot be detected directly if the old
compact stars are NSs.

However, the composition of pulsarlike compact stars remains unclear and
its quest is a big challenge in contemporary
science~\citep{XuRX03,Lat04,Alf07,Web09}.
Besides the conventional NSs, one important candidate is strange
quark stars (SQSs)~\citep{Ito70,Alf07,Web09}, made purely of deconfined {\it u}, {\it d}, and
{\it s} quark matter (with some leptons), i.e., strange quark matter (SQM)
which might be the true ground state of QCD matter and is absolutely stable
according to the Bodmer-Witten-Terazawa hypothesis~\citep{Wit84,Web09}.
Many probes
have been proposed to distinguish SQSs from NSs, e.g.,
SQSs have much larger dissipation rate of radial vibrations~\citep{Wang84}
and higher bulk viscosity~\citep{Haen89},
the spin rate of SQSs can be much closer to the Kepler limit than that
of NSs~\citep{Mad92}, SQSs may cool more rapidly than NSs within the
first $30$ years~\citep{Scha97}, the gravity-mode (g-mode)
eigenfrequencies in SQSs are much lower than those in
NSs~\citep{Liu08}, and so on.
In the present work, we show that
SQSs could be a good probe of the interactions
between light quarks and DM, and the observation of
scalar bosonic DM by future terrestrial experiments would favor old
pulsars to be SQSs rather than NSs.

This paper is organized as follows.
We first briefly introduce the methods of calculating the
accretion mass of DM by a compact star in Sec.~{\ref{sec:accretion}.
Then the compact star models are described in
Sec.~{\ref{sec:compactStar}}.
In Sec.~{\ref{sec:limits}},
we present the results and discussions.
Finally, the conclusions are given in Sec.~{\ref{sec:conclusion}.

\section{Accretion of DM}
\label{sec:accretion}
Following our recent work~\citep{Hao15}, we adopt a spherically symmetric
accretion scenario to calculate the capture rate of DM by a structured
compact star, and the total mass of DM captured by the star during a time period
of $t$ can be obtained as
\begin{eqnarray}
M_{t}
& = & 4.07 \times 10^{40} \, {\rm GeV} \, \frac{M_{S} R_{S}}{1 - 2.964 \frac{M_{S}}{R_{S}}} \left(\frac{m_\chi n_\chi}{0.3 \, {\rm GeV}/{\rm cm}^3}\right) \notag \\
& & \times \left( \frac{v_0}{220 \, {\rm km}/{\rm s}}\right)^{-1} \left(\frac{t}{{\rm Gyr}}\right) \, f \, ,
\label{total_mass}
\end{eqnarray}
where $M_S$ (in unit of solar mass $M_{\odot}$) and $R_S$ (in unit of \textit{km})
are the star's mass and radius, respectively; $m_{\chi}$ ($n_{\chi}$) is
the mass (number density) of DM particles which are assumed to follow
a Maxwell-Boltzmann distribution with the mode speed of $v_0$; $f$ is the
fraction of DM particles that undergo at least one collision
inside the star and it can be expressed as
\begin{equation}
f =
\left\langle 1 - e^{-\sum_i\int \sigma_{i} \xi_{i}(r) n_{i}(r) {\rm d}l} \right\rangle \, ,
\label{frac}
\end{equation}
with $\sigma_{i}$ denoting the scattering cross section between DM and star
constituent particle $i$ in free space,
$\xi_{i}(r)$ the medium
corrections due to the Pauli blocking effect and Fermi motion, and
the integration in the exponent is taken over the arc length along
the trajectory ($l$) of DM crossing through the star and the summation
is for various constituent particles of the star.
Finally, all the possible trajectories of DM inside the compact star
are averaged (denoted by the angle brackets) in Eq.~(\ref{frac}).
The DM-lepton interactions play
minor roles and are neglected in the calculations.

After being captured by a compact star, the DM
may become thermalized soon with the star matter and gather
at the host star center with a radius of about several
meters within a typical time period of $t_{\rm th} \sim 0.2 \, {\rm yr} \, \left(\frac{m_{\chi}}{{\rm TeV}}\right)^2 \left(\frac{\sigma_{b_i}}{{10^{-43}\, {\rm cm}^2}}\right)^{-1} \left(\frac{T_c}{T_5}\right)^{-1}$,
where $T_c$ is the central temperature of the star and
$T_5 = 10^5 \, {\rm K}$~\citep{Yu12,Ber13,Kumar13,Hao15}.
Especially, for bosonic DM, the BEC state, confined by the star's
gravitational field, can be formed when the number of
accumulated DM particles exceeds a critical
value of $N_{\rm BEC} \sim 2\times10^{35}(T_c/T_5)^3$.
The DM particles exceeding $N_{\rm BEC}$ will fall into
the ground state and gather within a tiny sphere with a
radius $r_{\rm BEC} \sim 1.4 \, m_{\chi}^{-1/2} \, {\rm \mu m}$,
where $m_{\chi}$ is in unit of GeV.
The DM in the BEC phase will quickly become self-gravitating
and form a boson star inside the host star.

As long as the boson star mass exceeds its Chandrasekhar limit,
i.e., $M_{t} > M_{\rm chan} + m_{\chi} N_{\rm BEC}$, it collapses
into a BH.
For non-interacting bosonic DM, we have $M_{\rm chan} = (2/\pi)M_{\rm pl}^2/m_{\chi}$
with the Planck mass $M_{\rm pl} = 1.22 \times 10^{19} \, {\rm GeV}$.
The newly born BH might grow up and eventually swallow the
host star if it accretes the star matter faster than
its evaporation through Hawking radiation (see, e.g., Eq.~(47) in Ref.~\citep{Hao15}).
Specifically, we adopt a spherically symmetric Bondi accretion
scenario~\citep{Sha83} to calculate the accretion rate of star matter,
and the BH is assumed to evaporate through Hawking
radiation of photons. The stable growth of BH mass $M_{\rm BH}$ is
guaranteed by the condition $\left.\frac{{\rm d} M_{\rm BH}}{{\rm d}t}\right|_{t=0} > 0$.
As a result,
the observation of old compact stars implies that either
$M_{\rm t} < M_{\rm chan} + m_{\chi} N_{\rm BEC}$
or $\left.\frac{{\rm d} M_{\rm BH}}{{\rm d}t}\right|_{t = 0} < 0$ must
be satisfied to prevent destruction of these stars. This means that the
interactions between DM and star matter cannot be too strong so as to prevent
the BH formation or stable growth of BH mass, which would put limits on the
scattering cross sections between DM and compact star matter.

\section{Compact star models}
\label{sec:compactStar}
In the present work, SQSs are assumed to be static and consist of neutrino-free SQM in
$\beta$-equilibrium with electric charge neutrality.
The equation of state (EOS) of SQM is taken from the MIT bag
model~\citep{Cho74} with two light flavors, i.e., the $u$ and $d$ quarks,
and one massive flavor, corresponding to the $s$ quark.
We further consider corrections in the thermodynamic potential
density due to perturbation theory to first order in $\alpha_s$
in the $\overline{{\rm MS}}$ scheme~\citep{Fra05,Kur10,Peng15}.
We select the QCD scale parameter $\Lambda_{\overline{\rm MS}}$ and
the invariant mass parameter $\hat{m}_s$ to be $146.2$ MeV and $279.9$ MeV,
respectively, according to Table.~1 in~\citep{Peng15}.
Then the QCD coupling and the $s$ quark mass are determined by
the running renormalization subtraction point $\Lambda$ given by
\begin{equation}
\Lambda = \frac{2}{3} (\mu_u + \mu_d + \mu_s) \, ,
\end{equation}
where $\mu_q$ ($q =u, d$ and $s$) denotes the chemical potential
of each flavor.
The bag constant has been chosen to be $B^{1/4} = 135.0 \, {\rm MeV}$ to
yield a value of $873.6$ MeV for the binding energy per baryon for cold SQM
in equilibrium versus a value of $954.7$ MeV for two-flavor $u$-$d$ quark
matter in equilibrium, and thus the absolutely stable condition is satisfied.
The SQS structure is then obtained by solving the
Tolman-Oppenheimer-Volkoff equations.
Shown in Fig.~\ref{composition} is the radial density distribution of
each quark flavor for a typical SQS with mass
$M_{S} = 1.4 \, M_{\odot}$, and the SQS radius is obtained as
$R_S = 11.47 \, {\rm km}$. Based on Fig.~\ref{composition},
the $f$ in Eq.~(\ref{frac}) can then be evaluated.
We note that using the MIT bag
model without perturbation corrections for the quark interactions~\citep{Liu08}
does not change our conclusions.

For NSs, we adopt the conventional neutron star model
in which the NS is assumed to consist of $\beta$-stable and
electrically neutral $npe\mu$ matter and the EOS
is taken from Skyrme-Hartree-Fock approach using the MSL1
interaction~\citep{Zhang13}, and the details can be found in Ref.~\citep{Hao15}.

\begin{figure}[!htb]
\includegraphics[width=7.5 cm]{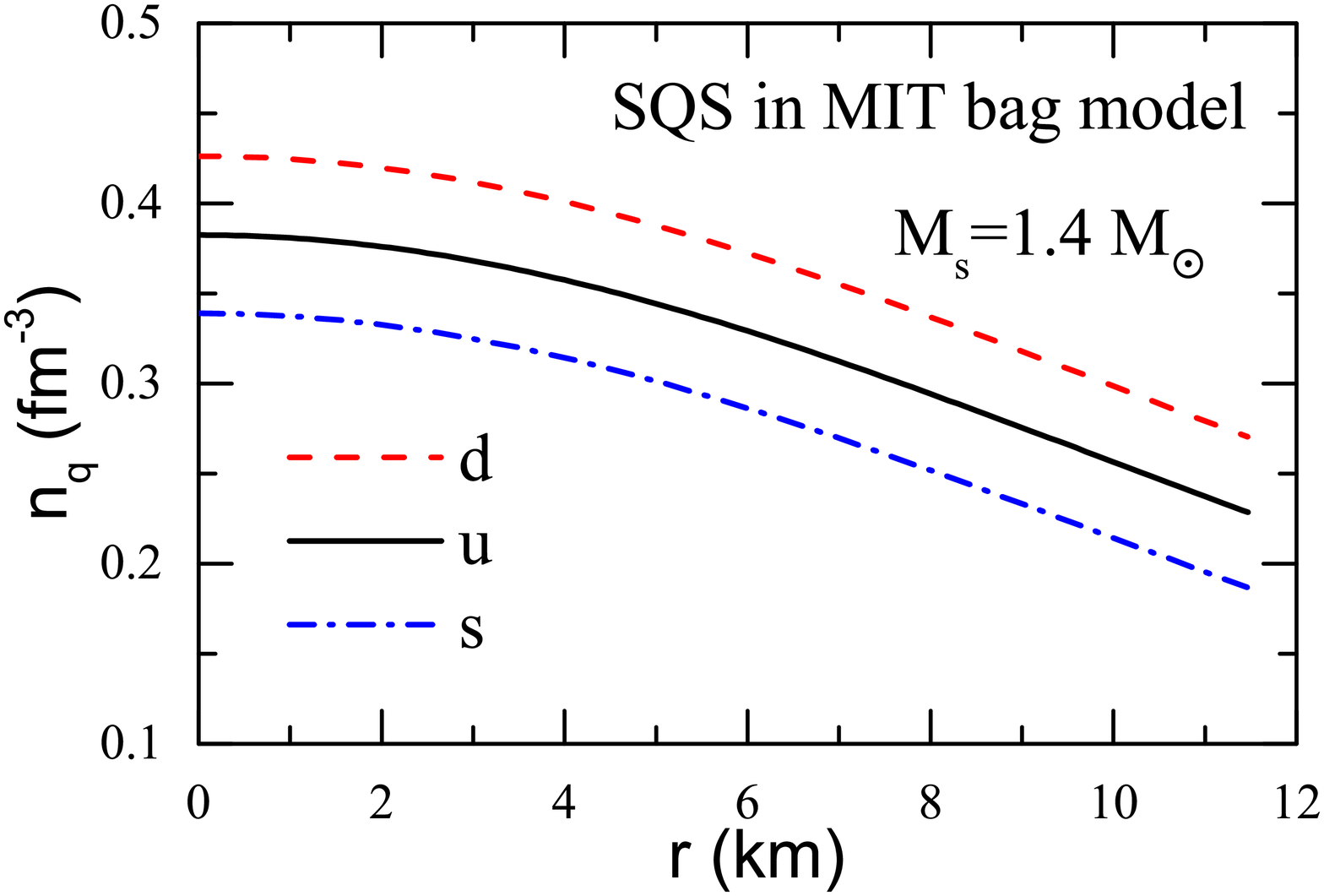}
\caption{(Color online)
Radial quark number density distribution for a SQS with mass of $1.4 \, M_{\odot}$
from the MIT bag model.}
\label{composition}
\end{figure}

The thermalization and BEC formation of DM in the interior of compact stars
depend on significantly the star's internal temperature $T_c$ which is
largely uncertain in observations.
For NSs, following our previous work~\citep{Hao15},
the $T_c$ can be estimated from the relatively well-studied NS's surface
temperature $T_s$ based on the study of thermal structure of the insulating
envelope presented by~\cite{Gud82}.
In particular, \cite{Gud82} found that the temperature $T_b$ measured at inner
boundary of the envelope (with density $\rho \sim 10^{10} \, {\rm g \, cm^{-3}}$)
can be related to the surface temperature $T_s$ as
\begin{equation}
T_{b} = 1.288\times10^8 \, {\rm K} \left[\frac{10^{14} \, {\rm cm/s^2}}{g_{s}}
\left(\frac{T_{s}}{10^6 \, {\rm K}}\right)^4\right]^{0.455} \, ,
\label{Tc}
\end{equation}
where $g_{s}$ is the surface gravity.
Since old compact stars are already isothermal in their interiors,
the $T_{c}$ can be assumed to be equal to $T_{b}$.

For SQSs, contrary to a bare surface with a steep density drop
over a few \textit{fm}, they may be wrapped up either in a tiny crust consisting
of ``normal" matter, i.e., ions and electrons, with maximum density
below the neutron drip density~\citep{Alc86}, or in a heterogeneous
(solid) crust made of strange nuggets and electrons~\citep{Jai06}.
The former crust is supported by strong electric fields near the surface and
has similar properties with the thermal envelope in Gudmundsson's model.
Thus it plays the role of an insulating layer and in this case we assume
the same relationship between $T_{c}$ and $T_{s}$ as shown in Eq.~(\ref{Tc}),
which leads to an estimate of $T_{c}/T_{s} \simeq 15.1 \times (T_{s}/T_5)^{0.82}$
based on the SQS structure shown in Fig.~\ref{composition}.
By assuming $T_{s} \sim T_5$, our estimate is nicely consistent with previous
assumption presented by~\cite{Bla00} and~\cite{Hor91} with $T_{c}/T_{s} = 20$ in their work.
For the heterogeneous crust within the strange nuggets model, \cite{Jai06}
found a relatively large crust with radial extent $\Delta R \simeq 40 \, {\rm m}$
and maximum density up to $\rho \sim 10^{13} \, {\rm g \, cm^{-3}}$.
The opacity of the nuggets matter has two origins, one due to scattering of
electrons off nuggets, and the other due to scattering among
different electrons.
Since the nuggets phase resembles the mixed phase of nuclei and electrons in
the crust of normal neutron stars~\citep{Jai06,Page04}, thus we can also
apply Eq.~(\ref{Tc}) to estimate $T_{c}$ from $T_{s}$.
It should be noted that crust consisting of strange nuggets may exist only if
the surface tension between the zero-pressure surface and the vacuum satisfies
condition $\sigma < \sigma_{\rm crit}$~\citep{Jai06}.
For the MIT bag model in this work, we have
$\sigma \approx 4.1 \, {\rm MeV \cdot fm^{-2}}$ and
$\sigma_{\rm crit} \approx 135 \, {\rm MeV \cdot fm^{-2}}$, implying
that a crust of mixed phase is indeed favored in our model.
The above discussions suggest that Eq.~(\ref{Tc}) provides a reasonable
approach to estimate $T_c$ from $T_s$ for both NSs and SQSs.

Furthermore, for solitary compact stars older than several million years,
the $T_{s}$ is estimated to be lower than $10^{5} \, {\rm K}$~\citep{Ya04,Neg12}.
In the present work, we assume a fixed $T_{s} = 10^{5} \, {\rm K}$ for
both NSs and SQSs. Since the DM particles accumulated in old compact stars (e.g.,
those with age older than $10^9$ years that we are interested in) essentially
come from the accretion process at the late stage of thermal evolution when
the $T_s$ is lower than $10^{5}$ K,
the resulting constraints in the present work are thus expected to be
conservative as a lower $T_{s}$ will lead to stronger constraints.

\section{Limits on $\sigma_q$ and $\sigma_p$}
\label{sec:limits}
Since numerous pulsarlike compact stars have been observed in
the Milky Way with various states, the resulting constraints
from different stars on DM-quark (proton) spin-independent (SI)
scattering cross sections $\sigma_q$ ($\sigma_p$)
should vary from one star to another.
We thus scan over all the available solitary compact objects and
finally figure out the one leading to the most stringent limits.
Particularly,
we only focus on the isolated pulsar systems to
avoid the additional complexity due to the evolution history of
pulsars in a binary (or more complex) system.
Moreover, the pulsars with age less than $1$ million years are ignored
since they are expected to have little time to accrete enough
DM and to have a relatively higher temperature than
the older ones.
Whereas there are little information on the masses and radii of the solitary
pulsars, we assume all of them have the fiducial mass of
$1.4 \, M_{\odot}$ with the radii calculated from the NS or SQS
models.
From Eq.~(\ref{total_mass}), for compact stars with the same
structure, the variation of the constraints from different
compact stars is mainly due to the term
$w_{\chi}(r) = \rho_{\chi}(r) \cdot t$.
Here the living age $t$ is taken as the pulsar's spin-down age.
The DM mass density $\rho_{\chi}(r) = m_{\chi}n_{\chi}(r)$ depends on the
halo model for which we adopt here the spherically symmetric generalized
Navarro-Frenk-White (NFW) profile~\citep{NFW} and Einasto
profile~\citep{Einasto}, i.e.,
\begin{equation}
\rho_{\chi}(r) =
\begin{cases}
\frac{\bar{\rho}_{s}}{\left(1 + \frac{r}{r_{s}}\right)^{3-\alpha} \left(\frac{r}{r_{s}}\right)^{\alpha}} & \text{(NFW)}, \\
\bar{\rho}_{s} \, e^{-\frac{2}{\alpha} \left[ \left(\frac{r}{r_{s}}\right)^{\alpha} - 1 \right]} & \text{(Einasto)},
\end{cases}
\label{halo_profile}
\end{equation}
where $r_{s}$ is the scale radius, $\bar{\rho}_{s}$ is the scale density and
$\alpha$ is the inner slope for the NFW profile and a shape parameter for the
Einasto profile.
We take $\alpha = 1$ (0.17) for NFW (Einasto) based on the results of N-body
simulations~\citep{Nava10} and $r_{s} = 20 \, {\rm kpc}$~\citep{Iocco11}.
The $\bar{\rho}_{s}$ is then obtained by fitting
the solar system DM density $\rho_{0} = 0.4 \, {\rm GeV/cm^3}$.
We scan over
all the available $1403$ solitary pulsars recorded in the ATNF Pulsar
Catalogue~\citep{Man05} and find the one maximizing $w_{\chi}(r)$ is
PSR J1801-0857D, a solitary pulsar with age of $9.71 \, {\rm Gyr}$ and
distance of $3.06 \, {\rm kpc}$ from the galactic center.
The corresponding
$w_{\chi}(r)$ is $16.0 \, (19.1) \, {\rm GeV \cdot Gyr \cdot cm^{-3}}$
for NFW (Einasto), indicating small model
dependence on the halo profiles for our results.
It should be emphasized that, within the present framework,
PSR J1801-0857D can set the strongest constraints on DM-quark and DM-proton
scattering cross sections among all the $1403$ pulsars. All results
in the following are calculated by using parameters of PSR J1801-0857D.

Now we can directly constrain $\sigma_{u}$ ($\sigma_{p}$) from the
existence of PSR J1801-0857D by assuming it is a SQS (NS).
Furthermore, one can convert the limits on $\sigma_{u}$ obtained from the
SQS assumption to those on $\sigma_{p}$, and then compare with the
constraints obtained from the NS assumption as well as the results
released by various direct detection experiments.
Based on general operator analyses, for scalar DM, the
effective operators describing the DM-quark interactions that can generate
DM-nucleon SI scattering are limited to the following two
classes~\citep{Gao13}:
\begin{equation}
\begin{split}
a_{q} \, \phi^{\dag}\phi \, \bar{q}q , \;\;\;\;
b_{q} \, \phi^{\dag}\overleftrightarrow{\partial}^{\mu}\phi \, \bar{q}\gamma_{\mu}q \, .
\label{effective_operator}
\end{split}
\end{equation}
The above first and second operators lead to
the scalar and vector DM-quark interactions, respectively, with
$a_{q}$ and $b_{q}$ being the coupling coefficients.
The corresponding effective operators describing the DM-nucleon
interactions are $f_{N}^{s} \, \phi^{\dag}\phi \, \bar{N}N$ and
$f_{N}^{v} \, \phi^{\dag}\overleftrightarrow{\partial}^{\mu}\phi \, \bar{N}\gamma_{\mu}N$,
respectively, where $N$ denotes protons ({\it p}) or neutrons ({\it n})
and $f_{N}$ is related to $a_{q}$ and $b_{q}$ by
\begin{equation}
f_{N}^{s} = \frac{1}{2 m_{\chi}} \sum_{q} B_{q}^{N} a_{q} \, , \;\;
f_{N}^{v} = \sum_{q} B_{q}^{N} b_{q}
\label{eff_oper_relation}
\end{equation}
with the dimensionless quantities~\citep{Gao13}
$B_{u}^{p} = 9.3$, $B_{u}^{n} = 6.5$, $B_{d}^{p} = 5.1$, $B_{d}^{n} = 7.1$
and $B_{s}^{p,n} = 1.2$ for scalar interaction,
and $B_{u}^{p} = 2$, $B_{u}^{n} = 1$, $B_{d}^{p} = 1$, $B_{d}^{n} = 2$
and $B_{s}^{p,n} = 0$ for vector interaction.
Eq.~(\ref{eff_oper_relation}) does not include the contributions of
sea quarks and gluons to $f_{N}^{s}$, which can be effectively
encoded in an additional free coefficient~\citep{Ciri14} and will be
discussed later.
Then the DM-proton scattering cross section, for both scalar and vector
interactions, takes the form $\sigma_{p} = (\mu_{p}^2/\pi)f_{p}^2$ while the
DM-quark scattering cross sections are given by $\sigma_{q}^{s} = (1/4\pi)(\mu_{q}^2 a_{q}^2/m_{\chi}^2)$
for scalar interaction and $\sigma_{q}^{v} = \mu_{q}^2 b_{q}^2/\pi$ for
vector interaction.
Here $\mu_{p}$ ($\mu_{q}$) is the DM-proton (DM-quark) reduced mass.
In the present work, the current masses of various quark flavors
are taken as $m_{u} = 2.3 \, {\rm MeV}$, $m_{d} = 4.8 \, {\rm MeV}$,
and $m_{s} = 95 \, {\rm MeV}$.
Thus one can derive limits on $\sigma_{p}$ from those on
$\sigma_{q}$ with a specific type of DM-quark interaction.

\begin{figure}[!htb]
\includegraphics[width=8.5 cm]{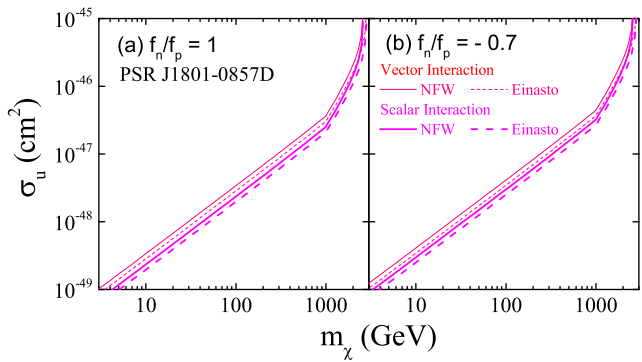}
\caption{(Color online)
Limits in the $m_{\chi}$-$\sigma_{u}$ plane for scalar DM by assuming
PSR J1801-0857D to be a SQS. DM is assumed to only interact with the first family of
quarks.}
\label{limits_ccs_01}
\end{figure}

Shown in Fig.~{\ref{limits_ccs_01}} are the limits on $\sigma_{u}$ vs $m_{\chi}$ by assuming that
DM only interact with the first family of quarks for two cases of the so-called
isospin-violating DM~\citep{Feng11} with $f_{n}/f_{p} = 1$
and $-0.7$.
Note that the cut-off mass around several TeV in Fig.~\ref{limits_ccs_01}
implies that PSR J1801-0857D fails to put constraints on heavier DM since for them
Hawking radiation will always overwhelm the
BH accretion and so the BH cannot grow up stably.
In addition, $B_{u}^{N}$ and $B_{d}^{N}$
are different for scalar and vector interactions,
and so the limits show an
interaction dependence with a stronger limit from the scalar interaction.
On the other hand,
the limits with NFW and Einasto only display a very small difference as expected.
From Fig.~{\ref{limits_ccs_01}}, one can see that PSR J1801-0857D
indeed can put extremely strong limits on $\sigma_{u}$, especially for light DM, at the
order of $10^{-49}$ cm$^2$. Since similar results on $\sigma_{d}$
can be obtained via relationship
$\sigma_{d}/\sigma_{u} = (\mu_{d}/\mu_{u})^{2} g_{du}^{2}$ where
$g_{du} = a_{d}/a_{u} \; (b_{d}/b_{u})$ for scalar (vector) interaction is related to the isospin-violating
factor $g_{np} = f_{n}/f_{p}$
by $g_{du} = (g_{np}B_{u}^{p} - B_{u}^{n}) / (B_{d}^{n} - g_{np}B_{d}^{p})$, our
present results are potentially useful in constraining
various model parameters for
DM-quark interactions~\citep{Gao13}.

\begin{figure}[!htb]
\includegraphics[width=8.5 cm]{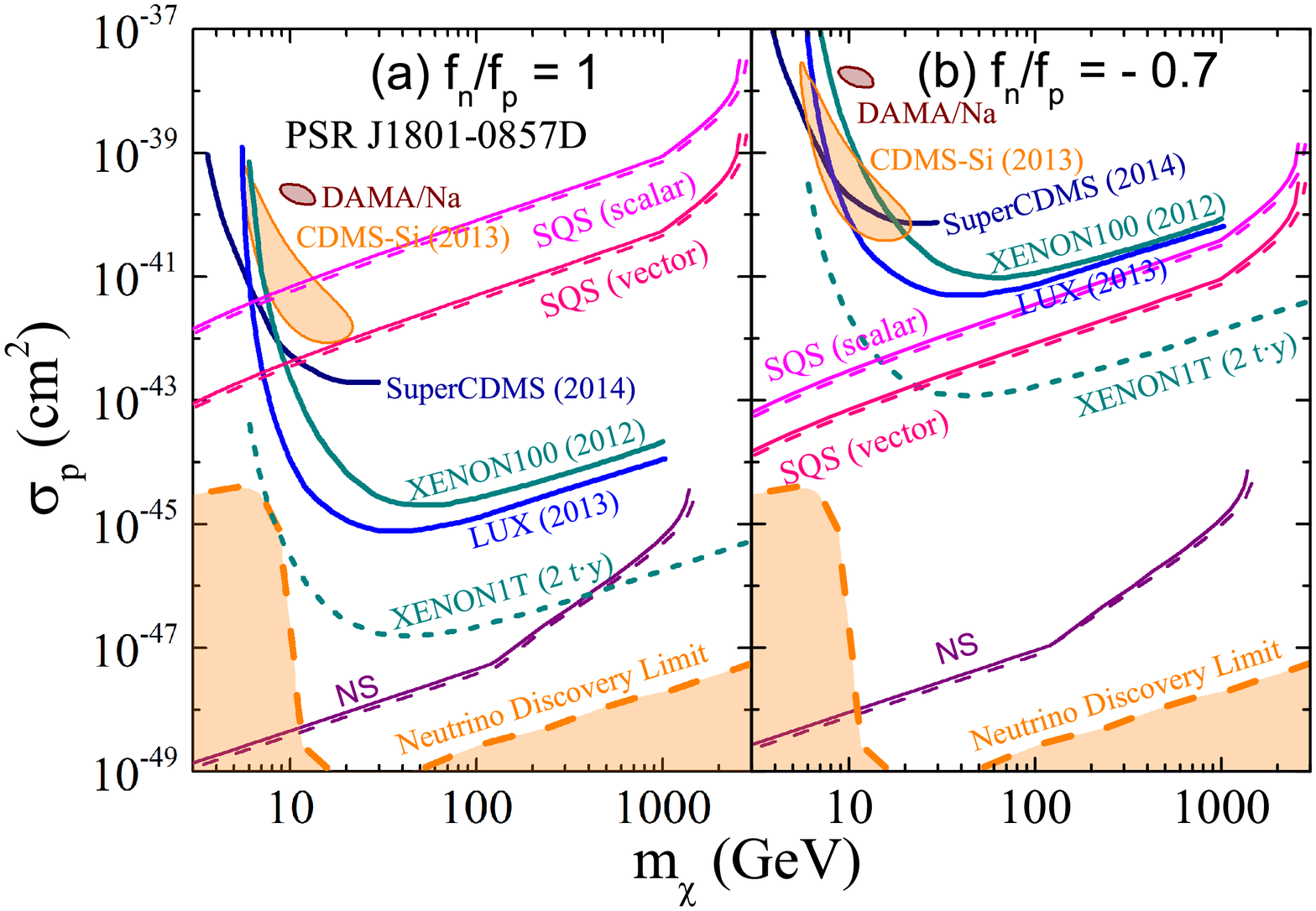}
\caption{(Color online)
Limits in the $m_{\chi}$-$\sigma_{p}$ plane for scalar DM by assuming
PSR J1801-0857D to be a SQS or a NS with NFW (solid lines) and Einasto (dashed lines)
DM halo profile. For the SQS case, DM is assumed to
only interact with the first family of quarks. The corresponding results
from direct detection experiments, i.e., DAMA-Libra~\citep{DAMA08},
CDMS-II(Si)~\citep{CDMS13}, XENON100~\citep{Xe12},
LUX~\citep{Lux13}, SuperCDMS(Ge)~\citep{SCDMS14}
as well as the future XENON1T~\citep{Xe1t}
and the ``neutrino discovery limit"~\citep{Bill14} are also
included for comparison.}
\label{limits_ccs_02}
\end{figure}

Then we move to the $m_{\chi}$-$\sigma_{p}$ plane.
Shown in Fig.~{\ref{limits_ccs_02}}
are the constraints on $\sigma_{p}$ either obtained directly
from the NS assumption or derived from the $\sigma_{u}$
as shown in Fig.~{\ref{limits_ccs_01}} from the SQS assumption.
It is very interesting to see that,
for both cases of $f_{n}/f_{p} = 1$ and $-0.7$, the limits on
$\sigma_{p}$ derived from $\sigma_{u}$ are shifted upward dramatically
compared with that on $\sigma_{u}$
and become significantly larger than the limits given by NS.
This is due to the fact that the $\sigma_{p}$ converted from the $\sigma_{q}$
is enlarged by a factor of $(\mu_{p}/\mu_{q})^2 (B_{u}^{p}+g_{du}B_{d}^{p})^2$
with an amplitude of about $10^6 - 10^7$,
and thus the limits on $\sigma_{p}$ set by the SQS assumption are significantly weakened
compared with that set by the NS assumption.
For comparison, the current limits and regions on $\sigma_{p}$ reported by
various direct detection experiments are also shown in Fig.~{\ref{limits_ccs_02}},
including the regions from
DAMA-Libra~\citep{DAMA08} and CDMS-Si~\citep{CDMS13} experiments and the limits
from SuperCDMS~\citep{SCDMS14}, XENON100~\citep{Xe12} and LUX~\citep{Lux13}
groups.
Also included are the expected sensitivity of future experiment XENON1T~\citep{Xe1t}
and the ``neutrino discovery limit"~\citep{Bill14} which sets limit on the
sensitivity of the direct detection method.

It is seen from Fig.~{\ref{limits_ccs_02}} that, for both cases with $f_{n}/f_{p}=1$
and $-0.7$,
the NS assumption provides the most stringent constraints.
In particular, even for the isospin-invariant case that $f_{n}/f_{p}=1$,
they are beyond the sensitivity of future XENON1T up to $m_{\chi}\sim 400 \, {\rm GeV}$.
However, the constraints
set by SQS assumption with scalar interaction become compatible with the
CDMS-Si contour for $f_{n}=f_{p}$, and the relatively
more stringent constraints from vector interaction are still weaker
than that of the current xenon-based experiments.
Moreover,
for the isospin-violating case with $f_{n}/f_{p} = -0.7$ (i.e.,
the so-called xenophobic DM), while
the tension among various direct detection experiments is largely ameliorated due to
the destructive interference of DM scattering with protons and neutrons
inside the target nuclei, all the currently favored DM regions are
excluded even by the softest limits from the SQS assumption.
But it is interesting to see that
future XENON1T, even in the xenophobic case,
is expected to have higher sensitivity than
the limits set by the SQS assumption for massive DM.
This means that if
future experiments had observed scalar DM signals within the mass region
$\gtrsim \mathcal{O}(10) \, {\rm GeV}$,
then the old compact objects are favored to be SQSs rather than NSs.
On the other hand, if no positive signals were observed by
XENON1T, the old compact stars still provide important
constraints on models which predict
scalar DM with mass lighter than $\sim \mathcal{O}(10) \, {\rm GeV}$.
Therefore direct detection experiments for DM
provide a novel way to probe the nature of pulsars.

\begin{figure}[!htb]
\includegraphics[width=8.5 cm]{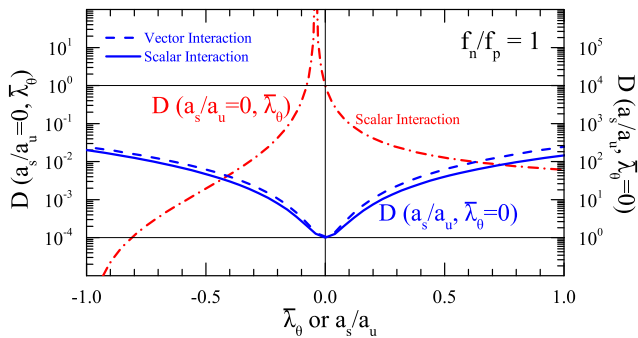}
\caption{(Color online)
Degradation factor due to the contributions from $s$ quark and heavy
quarks (gluon) for the limits in the $m_{\chi}$-$\sigma_{p}$ plane
by assuming PSR J1801-0857D to be a SQS in the case of $f_{n}=f_{p}$.}
\label{degradation}
\end{figure}

Finally, we discuss the effects of $s$ quark and heavy quarks (gluon)
by showing in Fig.~\ref{degradation} the so-called
degradation factor~\citep{Feng13} which measures
the suppression or amplification effects on the $\sigma_{p}$ constraints
and is defined as
\begin{equation}
D(g_{su}, \overline{\lambda}_{\theta}) = \frac{\sigma_{p}(0, 0)}{\sigma_{p}(g_{su}, \overline{\lambda}_{\theta})} \, ,
\end{equation}
with $g_{su} = a_{s}/a_{u}$ and $\overline{\lambda}_{\theta}$
the rescaled heavy quark (gluonic) coupling defined in Ref.~\citep{Ciri14}.
Noting that the isospin-violating effect is very small for large
$g_{su}$ or $\overline{\lambda}_{\theta}$, we thus only show the results
with $f_{n}/f_{p} = 1$ in Fig.~\ref{degradation}.
It is interesting to see $D(g_{su}, 0)\geq 1$ for all $g_{su}$, and
thus including $s$ quark contributions
can significantly increase the sensitivity of the $\sigma_{p}$ constraints
set by a SQS for both scalar and vector DM-quark interactions.
On the other hand, for the heavy quark (gluon) effects with
vector interaction, the sea quarks and gluons do not contribute
to $f_{p,n}$ due to the conservation of vector current, and
thus we show the degradation factor only for
scalar interaction in Fig.~\ref{degradation}.
One can see the heavy quark (gluon) effects
can either increase or decrease the sensitivity,
depending on the specific value of $\overline{\lambda}_{\theta}$.
To further constrain $g_{su}$ and $\overline{\lambda}_{\theta}$,
other independent constraints are necessary, e.g., from
the collider experiments~\citep{Goo10}.

\section{Conclusions}
\label{sec:conclusion}
For scalar DM with ignorable self-interactions, we have
shown that the old strange quark stars can directly put important
constraints on DM-quark scattering cross sections $\sigma_{q}$, which
can be further converted into the constraints on DM-proton scattering
cross sections $\sigma_{p}$ based on effective operator analyses.
By analyzing a set of 1403 of
solitary pulsarlike compact stars in the Milky Way,
we have found that the old pulsar PSR J1801-0857D can put the most
stringent constraints on $\sigma_{q}$ and $\sigma_{p}$.
Furthermore, we have demonstrated that while
the limits on $\sigma_{p}$ obtained by assuming PSR J1801-0857D to be a neutron star
essentially rule out the
possibility of detecting scalar DM in terrestrial
labs through direct detection experiments, the extracted limits from
assuming PSR J1801-0857D to be a strange quark star are significantly
weakened to be comparable with the terrestrial direct detection experiments.
Our results have indicated that DM direct detection experiments provide
a novel way to probe the nature of old pulsarlike compact stars, and the old
pulsars are favored to be strange quark stars rather than neutron stars if
scalar DM could be observed by future direct detection experiments,
e.g., XENON1T.

\begin{acknowledgments}
We are grateful to Ren-Xin Xu for helpful discussions.
This work was supported in part by the Major State Basic
Research Development Program (973 Program) in China under Contract Nos.
2015CB856904 and 2013CB834405, the NSFC under Grant Nos. 11275125
and 11135011, the ``Shu Guang" project supported by Shanghai Municipal
Education Commission and Shanghai Education Development
Foundation, the Program for Professor of Special Appointment (Eastern Scholar)
at Shanghai Institutions of Higher Learning, and the Science and Technology
Commission of Shanghai Municipality (11DZ2260700).
\end{acknowledgments}

%

\end{document}